\documentclass[twocolumn]{aastex62}
\newcommand{\thisstar}{GJ 1214}

\usepackage{amsmath}
\usepackage{graphicx}
\usepackage{natbib}
\usepackage[scaled]{helvet}
\usepackage{epsfig}
\usepackage{url}
\usepackage{textcomp}
\usepackage{mathpazo}
\usepackage{xspace}
\usepackage{hyperref}
\usepackage{apjfonts}
\usepackage{mathrsfs}
\usepackage{verbatim}
\usepackage{color}
\usepackage[normalem]{ulem}
\usepackage{lineno}

\bibpunct{(}{)}{;}{a}{}{,}
\newcommand{\gjeup}{\ensuremath{<0.019}}
\newcommand{\gjeupfac}{\ensuremath{3.32}}

\newcommand{\gjmstar}{\ensuremath{0.1820^{+0.0042}_{-0.0041}}}

\newcommand{\gjmstarfac}{\ensuremath{2.41}}
\newcommand{\gjrstar}{\ensuremath{0.2162^{+0.0025}_{-0.0024}}}
\newcommand{\gjrstarerr}{\ensuremath{1.13}}

\newcommand{\gjrstarfac}{\ensuremath{3.27}}
\newcommand{\gjrstarsed}{\ensuremath{0.2207^{+0.0030}_{-0.0033}}}

\newcommand{\gjlstar}{\ensuremath{0.00390^{+0.00021}_{-0.00020}}}

\newcommand{\gjlstarfac}{\ensuremath{3.17}}
\newcommand{\gjfbol}{\ensuremath{5.82^{+0.31}_{-0.30}}$\times 10^{-10}$}
\newcommand{\gjfbolerr}{\ensuremath{5.24}}
\newcommand{\gjrhostar}{\ensuremath{25.41^{+0.62}_{-0.71}}}
\newcommand{\gjrhostarerr}{\ensuremath{2.62}}

\newcommand{\gjrhostarfac}{\ensuremath{4.89}}
\newcommand{\gjlogg}{\ensuremath{5.0286^{+0.0078}_{-0.0088}}}

\newcommand{\gjloggfac}{\ensuremath{4.82}}
\newcommand{\gjteff}{\ensuremath{3101\pm43}}
\newcommand{\gjtefferr}{\ensuremath{1.39}}
\newcommand{\gjteffsigdiff}{\ensuremath{-1.37}}
\newcommand{\gjtefffac}{\ensuremath{2.33}}
\newcommand{\gjteffsed}{\ensuremath{3077^{+39}_{-35}}}

\newcommand{\gjfeh}{\ensuremath{0.24\pm0.11}}

\newcommand{\gjfehfac}{\ensuremath{1.09}}
\newcommand{\gjav}{\ensuremath{0.22^{+0.15}_{-0.14}}}

\newcommand{\gjsigmased}{\ensuremath{2.03^{+0.73}_{-0.47}}}

\newcommand{\gjksabs}{\ensuremath{7.953\pm0.020}}

\newcommand{\gjksabsfac}{\ensuremath{1.00}}
\newcommand{\gjksapp}{\ensuremath{8.781\pm0.020}}

\newcommand{\gjparallax}{\ensuremath{68.298^{+0.085}_{-0.084}}}

\newcommand{\gjparallaxfac}{\ensuremath{0.66}}
\newcommand{\gjdistance}{\ensuremath{14.642\pm0.018}}
\newcommand{\gjdistanceerr}{\ensuremath{0.12}}

\newcommand{\gjdistancefac}{\ensuremath{0.67}}
\newcommand{\gjperiod}{\ensuremath{1.580404531^{+0.000000018}_{-0.000000017}}}

\newcommand{\gjperiodsigdiff}{\ensuremath{1.53}}
\newcommand{\gjperiodfac}{\ensuremath{7.43}}
\newcommand{\gjrp}{\ensuremath{2.733^{+0.033}_{-0.031}}}

\newcommand{\gjrpfac}{\ensuremath{1.59}}
\newcommand{\gjmp}{\ensuremath{8.41^{+0.36}_{-0.35}}}

\newcommand{\gjmpfac}{\ensuremath{1.21}}
\newcommand{\gjtc}{\ensuremath{2459639.7812619^{+0.0000090}_{-0.0000089}}}

\newcommand{\gjtcfac}{\ensuremath{36.87}}
\newcommand{\gjtt}{\ensuremath{2459636.6204528^{+0.0000090}_{-0.0000089}}}

\newcommand{\gja}{\ensuremath{0.01505\pm0.00011}}

\newcommand{\gjafac}{\ensuremath{2.36}}
\newcommand{\gji}{\ensuremath{88.980^{+0.094}_{-0.085}}}

\newcommand{\gjisigdiff}{\ensuremath{2.13}}
\newcommand{\gjifac}{\ensuremath{1.12}}
\newcommand{\gje}{\ensuremath{0.0062^{+0.0079}_{-0.0044}}}

\newcommand{\gjomega}{\ensuremath{77.0^{+7.5}_{-61}}}

\newcommand{\gjteq}{\ensuremath{567.0\pm7.6}}

\newcommand{\gjteqfac}{\ensuremath{2.50}}
\newcommand{\gjtcirc}{\ensuremath{0.0218\pm0.0014}} 

\newcommand{\gjk}{\ensuremath{14.38^{+0.57}_{-0.56}}}

\newcommand{\gjkfac}{\ensuremath{0.94}}
\newcommand{\gjp}{\ensuremath{0.11589\pm0.00016}}

\newcommand{\gjpfac}{\ensuremath{1.06}}
\newcommand{\gjar}{\ensuremath{14.97^{+0.12}_{-0.14}}}

\newcommand{\gjarfac}{\ensuremath{1.23}}
\newcommand{\gjdelta}{\ensuremath{0.013430^{+0.000037}_{-0.000038}}}

\newcommand{\gjdeltasigdiff}{\ensuremath{-3.76}}
\newcommand{\gjdeltafac}{\ensuremath{1.07}}
\newcommand{\gjdeltaI}{\ensuremath{0.0173^{+0.0012}_{-0.0011}}}

\newcommand{\gjdeltaMIRI}{\ensuremath{0.01382\pm0.00021}}

\newcommand{\gjdeltaV}{\ensuremath{0.0161^{+0.0021}_{-0.0016}}}

\newcommand{\gjtau}{\ensuremath{0.004023^{+0.000050}_{-0.000051}}}

\newcommand{\gjtonefour}{\ensuremath{0.036236^{+0.000046}_{-0.000045}}}

\newcommand{\gjtfwhm}{\ensuremath{0.032212^{+0.000043}_{-0.000042}}}

\newcommand{\gjb}{\ensuremath{0.264^{+0.020}_{-0.023}}}

\newcommand{\gjbfac}{\ensuremath{1.16}}
\newcommand{\gjbs}{\ensuremath{0.268^{+0.021}_{-0.023}}}

\newcommand{\gjtaus}{\ensuremath{0.004078^{+0.000087}_{-0.000072}}}

\newcommand{\gjtonefours}{\ensuremath{0.03664^{+0.00055}_{-0.00038}}}

\newcommand{\gjtfwhms}{\ensuremath{0.03257^{+0.00048}_{-0.00033}}}

\newcommand{\gjdeltatwofive}{\ensuremath{2.83^{+0.32}_{-0.29}}}

\newcommand{\gjdeltafivezero}{\ensuremath{129.1^{+6.4}_{-6.3}}}

\newcommand{\gjdeltasevenfive}{\ensuremath{403\pm12}}

\newcommand{\gjrhop}{\ensuremath{2.26\pm0.11}}

\newcommand{\gjrhopfac}{\ensuremath{1.50}}
\newcommand{\gjloggp}{\ensuremath{3.043\pm0.019}}

\newcommand{\gjloggpfac}{\ensuremath{1.47}}
\newcommand{\gjsafronov}{\ensuremath{0.01793^{+0.00074}_{-0.00073}}}

\newcommand{\gjinsol}{\ensuremath{0.0234^{+0.0013}_{-0.0012}}}

\newcommand{\gjinsolfac}{\ensuremath{2.80}}
\newcommand{\gjtp}{\ensuremath{2459639.725^{+0.033}_{-0.27}}}

\newcommand{\gjts}{\ensuremath{2459638.99226\pm0.00020}}

\newcommand{\gjte}{\ensuremath{2459785.96988\pm0.00020}}

\newcommand{\gjta}{\ensuremath{2459640.9702^{+0.0040}_{-0.0028}}}

\newcommand{\gjtd}{\ensuremath{2459640.1740^{+0.0028}_{-0.0040}}}

\newcommand{\gjvcve}{\ensuremath{0.9940^{+0.0056}_{-0.0080}}}

\newcommand{\gjecosw}{\ensuremath{0.00137\pm0.00020}}

\newcommand{\gjesinw}{\ensuremath{0.0060^{+0.0080}_{-0.0056}}}

\newcommand{\gjmsini}{\ensuremath{8.41^{+0.36}_{-0.35}}}

\newcommand{\gjmsinifac}{\ensuremath{1.21}}
\newcommand{\gjq}{\ensuremath{0.0001389^{+0.0000056}_{-0.0000055}}}

\newcommand{\gjqfac}{\ensuremath{1.98}}
\newcommand{\gjdr}{\ensuremath{14.89^{+0.19}_{-0.25}}}

\newcommand{\gjpt}{\ensuremath{0.05939^{+0.0010}_{-0.00073}}}

\newcommand{\gjptg}{\ensuremath{0.07496^{+0.0013}_{-0.00093}}}

\newcommand{\gjps}{\ensuremath{0.05869^{+0.00031}_{-0.00032}}}

\newcommand{\gjpsg}{\ensuremath{0.07407^{+0.00041}_{-0.00042}}}

\newcommand{\gjuoneI}{\ensuremath{0.48^{+0.10}_{-0.11}}}

\newcommand{\gjutwoI}{\ensuremath{0.06\pm0.16}}

\newcommand{\gjuoneMIRI}{\ensuremath{0.062^{+0.029}_{-0.028}}}

\newcommand{\gjutwoMIRI}{\ensuremath{0.056\pm0.053}}

\newcommand{\gjuoneV}{\ensuremath{0.35^{+0.20}_{-0.19}}}

\newcommand{\gjutwoV}{\ensuremath{0.33^{+0.29}_{-0.30}}}

\newcommand{\gjdeltas}{\ensuremath{398\pm13}}

\newcommand{\gjgammapre}{\ensuremath{-3.19^{+0.55}_{-0.56}}}

\newcommand{\gjgammaprefac}{\ensuremath{2.98}}
\newcommand{\gjgammapost}{\ensuremath{-0.12^{+0.55}_{-0.58}}}

\newcommand{\gjgammapostfac}{\ensuremath{3.24}}
\newcommand{\gjsigmajpre}{\ensuremath{4.14^{+0.54}_{-0.48}}}

\newcommand{\gjsigmajprefac}{\ensuremath{1.10}}
\newcommand{\gjsigmajpost}{\ensuremath{3.29^{+0.63}_{-0.60}}}

\newcommand{\gjsigmajpostfac}{\ensuremath{1.15}}
\newcommand{\gjjittervarpre}{\ensuremath{17.1^{+4.8}_{-3.8}}}

\newcommand{\gjjittervarprefac}{\ensuremath{0.95}}
\newcommand{\gjjittervarpost}{\ensuremath{10.8^{+4.6}_{-3.6}}}

\newcommand{\gjjittervarpostfac}{\ensuremath{0.95}}

\begin{document}
\newcommand{\kgsins}[1]{\textcolor{blue}{#1}}
\newcommand{\kgsdel}[1]{\textcolor{red}{\sout{#1}}}
\newcommand\mynotes[1]{\textcolor{red}{#1}} 
\newcommand{\bjdtdb}{\ensuremath{\rm {BJD_{TDB}}}}
\newcommand{\tjdtdb}{\ensuremath{\rm {TJD_{TDB}}}}
\newcommand{\feh}{\ensuremath{\left[{\rm Fe}/{\rm H}\right]}}
\newcommand{\initfeh}{\ensuremath{\left[{\rm Fe}/{\rm H}\right]_0}}
\newcommand{\eep}{\ensuremath{\rm{EEP}}}
\newcommand{\mh}{\ensuremath{\left[{\rm M}/{\rm H}\right]}}
\newcommand{\teff}{\ensuremath{T_{\rm eff}}}
\newcommand{\teq}{\ensuremath{T_{\rm eq}}}
\newcommand{\fbol}{\ensuremath{F_{\rm bol}}}
\newcommand{\logk}{\ensuremath{\log K}}
\newcommand{\logg}{\ensuremath{\log g_*}}
\newcommand{\loggp}{\ensuremath{\log g_{\rm{P}}}}
\newcommand{\ecosw}{\ensuremath{e\cos{\omega_{*}}}}
\newcommand{\esinw}{\ensuremath{e\sin{\omega_{*}}}}
\newcommand{\lcosw}{\ensuremath{L\cos{\omega_{*}}}}
\newcommand{\lsinw}{\ensuremath{L\sin{\omega_{*}}}}
\newcommand{\vcve}{\ensuremath{V_c/V_e}}
\newcommand{\secosw}{\ensuremath{\sqrt{e}\cos{\omega_{*}}}}
\newcommand{\sesinw}{\ensuremath{\sqrt{e}\sin{\omega_{*}}}}
\newcommand{\msun}{\ensuremath{\,M_\odot}}
\newcommand{\rsun}{\ensuremath{\,R_\odot}}
\newcommand{\lsun}{\ensuremath{\,L_\odot}}
\newcommand{\lstar}{\ensuremath{\,L_*}}
\newcommand{\rhostar}{\ensuremath{\,\rho_*}}
\newcommand{\gstar}{\ensuremath{\,g_*}}
\newcommand{\obs}{\mathrm{obs}}
\newcommand{\tru}{\mathrm{true}}

\newcommand{\sigmasb}{\ensuremath{\,\sigma_{SB}}}

\newcommand{\mj}{\ensuremath{\,M_{\rm J}}}
\newcommand{\mearth}{\ensuremath{\,M_{\Earth}}}
\newcommand{\mplanet}{\ensuremath{\,M_{\rm P}}}
\newcommand{\rplanet}{\ensuremath{\,R_{\rm P}}}
\newcommand{\rhoplanet}{\ensuremath{\,\rho_{\rm P}}}
\newcommand{\msini}{\ensuremath{\,M_{\rm P}\sin{i}}}
\newcommand{\cosi}{\ensuremath{\cos{i}}}
\newcommand{\sini}{\ensuremath{\sin{i}}}
\newcommand{\rj}{\ensuremath{\,R_{\rm J}}}
\newcommand{\re}{\ensuremath{\,R_{\rm \Earth}}\xspace}
\newcommand{\me}{\ensuremath{\,M_{\rm \Earth}}\xspace}
\newcommand{\fave}{\ensuremath{\langle \rm F \rangle}}
\newcommand{\fluxcgs}{10$^9$ erg s$^{-1}$ cm$^{-2}$}
\newcommand{\newln}{\\&\quad\quad{}}
\newcommand{\kepler}{Kepler}
\newcommand{\ktwo}{K2}
\newcommand{\tess}{TESS}
\newcommand{\gaia}{{\it Gaia}}
\newcommand{\spitzer}{{\it Spitzer}}
\newcommand{\corot}{{\it CoRoT}}
\newcommand{\hipparcos}{{\it Hipparcos}}
\newcommand{\kmps}{\,km\,s$^{-1}$}
\newcommand{\ms}{\,m\,s$^{-1}$}
\newcommand{\minus}{\scalebox{0.75}[1.0]{$-$}}
\newcommand{\mstar}{\ensuremath{M_{*}}}
\newcommand{\rstar}{\ensuremath{R_{*}}}
\newcommand{\ar}{\ensuremath{a/R_*}}
\newcommand{\vsini}{\ensuremath{V\sin{I_*}}}
\newcommand{\exofasttwo}{{\tt EXOFASTv2}}
\newcommand{\exofast}{{\tt EXOFAST}}
\newcommand{\multifast}{{\tt MULTIFAST}}
\newcommand{\ks}{\ensuremath{K_{\rm S}}}
\newcommand{\tonefour}{\ensuremath{T_{14}}}
\newcommand{\tfwhm}{\ensuremath{T_{\rm FWHM}}}
\newcommand{\tfwhms}{\ensuremath{T_{\rm FWHM,S}}}
\newcommand{\tc}{\ensuremath{T_{\rm C}}}
\newcommand{\ts}{\ensuremath{T_{\rm S}}}
\newcommand{\tmps}{\ensuremath{T_{\rm T}}}
\newcommand{\tp}{\ensuremath{T_{\rm P}}}
\newcommand{\omegagr}{\ensuremath{\dot{\omega}_{\rm GR}}}

\newcommand{\Snom}{\hbox{$\mathcal{S}^{\rm N}_{\odot}$}}
\newcommand{\Rnom}{\hbox{$\mathcal{R}^{\rm N}_{\odot}$}}
\newcommand{\Mnom}{\hbox{$\mathcal{M}^{2014}_{\odot}$}}
\newcommand{\Lnom}{\hbox{$\mathcal{L}^{\rm N}_{\odot}$}}
\newcommand{\Tnom}{\hbox{$\mathcal{T}^{\rm N}_{\rm eff\odot}$}}
\newcommand{\Gnom}{\hbox{$\mathcal{G}^{\rm N}$}}
\newcommand{\GMnom}{\hbox{$\mathcal{(GM)}^{\rm N}_{\odot}$}}
\newcommand{\GMJnom}{\hbox{$\mathcal{(GM)}^{\rm N}_{\rm J}$}}
\newcommand{\GMEnom}{\hbox{$\mathcal{(GM)}^{\rm N}_{\rm E}$}}
\newcommand{\MJnom}{\hbox{$\mathcal{M}^{\rm N}_{\rm J}$}}
\newcommand{\MEnom}{\hbox{$\mathcal{M}^{\rm N}_{\rm E}$}}
\newcommand{\ReJnom}{\hbox{$\mathcal{R}^{\rm N}_{e\rm J}$}}
\newcommand{\RpJnom}{\hbox{$\mathcal{R}^{\rm N}_{p\rm J}$}}
\newcommand{\ReEnom}{\hbox{$\mathcal{R}^{\rm N}_{e\rm E}$}}
\newcommand{\RpEnom}{\hbox{$\mathcal{R}^{\rm N}_{p\rm E}$}}
\newcommand{\Mnew}{\mathcal{M}_{\odot}^{\rm new}}
\newcommand{\Mn}{\mathcal M_{\odot}^{2014}}
\newcommand{\MEn}{\mathcal M_{E}^{2014}}
\newcommand{\MJn}{\mathcal M_{J}^{2014}}
\newcommand{\Teff}{T_\mathrm{eff}}
\newcommand{\ls}{$L_{\odot}$}
\newcommand{\rs}{$R_{\odot}$}
\newcommand{\MnomS}{\hbox{$\mathcal{M}_{\odot}^{\rm N}$}}
\newcommand{\MnomJ}{\hbox{$\mathcal{M}_{\rm J}^{\rm N}$}}
\newcommand{\MnomE}{\hbox{$\mathcal{M}_{\rm E}^{\rm N}$}}
\newcommand{\um}{\ensuremath{\mu m}}

\title{Using JWST transits and occultations to determine $\sim$1\% stellar radii and temperatures of low-mass stars}

\author[0009-0000-8049-3797]{Alexandra S. Mahajan}
\affiliation{Center for Astrophysics \textbar \ Harvard \& Smithsonian, 60 Garden St, Cambridge, MA 02138, USA}

\author[0000-0003-3773-5142]{Jason D.\ Eastman}
\affiliation{Center for Astrophysics \textbar \ Harvard \& Smithsonian, 60 Garden St, Cambridge, MA 02138, USA}

\author[0000-0002-4207-6615]{James Kirk}
\affiliation{Department of Physics, Imperial College London, Prince Consort Road, London, SW7 2AZ, UK}

\correspondingauthor{Alexandra Mahajan}
\email{alexandra.mahajan@cfa.harvard.edu}

\shorttitle{GJ 1214}
\shortauthors{Mahajan, et. al.}

\begin{abstract}
Using JWST observations of a primary transit and two secondary eclipses for GJ 1214b, we determine an eccentricity that is more precise than a decade of HARPS data, which enables us to measure the stellar density to \gjrhostarerr\%. Coupled with a prior on the stellar mass from a dynamically calibrated \ks-\mstar \ relation, we determine \rstar \ to \gjrstarerr\%---3 times more precise than any other published analysis of this system. Then, using the bolometric flux from a spectral energy distribution model, we determine \teff \ to \gjtefferr\%---40\% more precise than systematic floors from spectroscopy. Within the global model, these also improve the planetary radius and insolation. This is a proof of concept for a new method to determine accurate \rstar \ and \teff \ to a precision currently achieved for only a small number of low-mass stars. By applying our method to all high signal-to-noise ratio planetary transits and occultations, we can expand the sample of precisely measured stars without assuming tidal circularization and calibrate new relations to improve our understanding of all low-mass stars.

\end{abstract}


\section{Introduction}
The masses, radii, and temperatures of planets depend directly on the same quantities as those of their host stars. Often, the limiting factor for determining precise planetary parameters is our knowledge of their host stars. Therefore, the higher the precision on the stellar parameters we can achieve, the better we will be able to understand climates, habitability, formation, and evolution of individual planets, as well as test and refine theoretical models for stellar atmospheres and evolution. This is the case for all planet-hosting stars, but is particularly important for low-mass host stars because they are the least well understood and they are  critical targets in the search for habitable Earth-like planets. 

The limitations of our understanding of stars are highlighted by \citet{Tayar:2022}, who showed that for spectroscopically measured temperatures, theoretical spectral energy distributions (SEDs), and evolutionary model constraints of stars, typical systematic errors are around 4.2\% in the stellar radius \rstar, 2.0\% in stellar temperature \teff, and 2.4\% in bolometric flux \fbol. It is widely accepted that the discrepancy between theory and observation gets larger for low-mass stars, where theoretically predicted stellar radii are only reliable to $\sim$10\% \citep[e.g.][]{Stauffer:2007,Bell:2012,Torres:2013,Thompson:2014}. 

While empirical relations improve these constraints, \citet{Casagrande:2014} noted significant discrepancies between measurements of \rstar \ from presumed gold-standard long-baseline optical interferometry (LBOI) using CHARA, and their own determination using the infrared flux method (IRFM). Despite the LBOI method's claimed precision of $\sim$1\% and the IRFM's precision of $\sim$3\% (limited by the $\sim$2\% systematic errors in the spectroscopic determination of \teff) the largest discrepancies were seen in the stars with the smallest angular size (regardless of physical size). This implies the gold-standard LBOI measurements were to blame. \citet{Tayar:2022} confirmed this behavior by highlighting discrepancies in different groups' LBOI measurements of the same systems using CHARA. More modern LBOI measurements (using additional calibration stars) may be more reliable, but still achieve a $\sim$1\% precision and only 29 stars are big, bright, and close enough to measure \citep{Mann:2015}. Another option for $\sim$1\% stellar radii is finding detached, double-lined eclipsing binaries, but these are also relatively rare, with only a handful $\mstar < 0.7 \msun$ known \citep{Torres:2010}.

Similar to IRFM, \citet{Mann:2015} determined a semiempirical relationship between the absolute \ks-band magnitude and the masses and radii of low-mass stars, where the masses and radii of the calibration sample were determined through LBOI measurements and spectroscopically determined temperatures and metallicities, with a surprisingly tight scatter: 1.8\% in mass and 2.7\% in radius. 

However, the discrepancy between the \mstar \ derived from the semiempirical \citet{Mann:2015} relations and the \mstar \ later determined from the empirical \ks-\mstar \ \citet{Mann:2019} relation calibrated dynamically exceeds the scatter of the relations in some cases (e.g., \thisstar) by a factor of 5. 

This implies strong systematic errors in the spectroscopic temperatures and metallicities used to calibrate the semiempirical models. That, and the fact that they also used the controversial LBOI radii to calibrate their relation, suggests the 2.7\% scatter about the \citet{Mann:2015} \ks-\rstar \ relation may not be a reliable estimate of its underlying uncertainty. That is, for most low-mass stars, the community optimistically achieves a radius precision of $\sim$3\%, limited by the $\sim$2\% systematic uncertainty in the spectroscopically determined \teff, and a mass precision of 2.3\%, limited by the scatter in the empirical \ks-\mstar \ relation \citep{Mann:2019}.

Meanwhile, \citet{Seager:2003} showed that Kepler's law can be rearranged in terms of the transit light curve's observables to directly measure the stellar density,

\begin{equation}
\label{eq:rearranged_rhostar}
    \rhostar = \frac{3\pi}{GP^2}\left(\frac{a}{R_*}\right)^3,
\end{equation}

\noindent where $G$ is the gravitational constant, $P$ is the planetary period constrained by the frequency of transits, and \ar, the planet's semi-major axis $a$ divided by \rstar, is constrained by the duration of the transit when the orbit is circular. Unfortunately, eccentricity $e$ and the argument of periastron $\omega_*$\footnote{Here we we mean the argument of periastron of the stellar orbit in a left-handed coordinate system, as is the century-old and commonly misunderstood convention. We have explicitly confirmed this definition when referencing equations stated ambiguously from other work.} change the planet's velocity during transit, making the transit duration more closely related to

\begin{equation}
    \label{eq:arobserved}
    \frac{a}{R_*}\frac{\sqrt{1-e^2}}{1+e\sin{\omega_*}}.
\end{equation}

That is, $\omega_*$, $e$, and \rhostar \ all change the observed transit duration (one observable and three unknowns). Typically, $e$ and $\omega_*$ are constrained from radial velocities, but to a precision that dominates the uncertainty in computing \rhostar \ from the light curve. However, with the duration of the primary transit, the duration of the secondary eclipse, and the phase offset between the two (three observables), we can precisely determine all three unknowns \citep{Kopal:1946,Charbonneau:2005,Ford:2008,Huber:2017,Alonso:2018,Eastman:2023} -- leading to an extremely precise determination of $e$ (and $\omega_*$ when $e\neq0$).
 
In particular, \citet{Alonso:2018} summarized the constraint on $e$ and $\omega_*$ from the transit and occultation as approximately

\begin{equation}
\label{eq:ecosw}
\ecosw = \frac{\pi}{P}\frac{\left(\ts-\tc-P/2\right)}{1+\csc ^2 i},
\end{equation} 

\noindent and

\begin{equation}
\label{eq:esinw}
\esinw = \frac{\beta\tfwhms -\tfwhm}{\beta \tfwhms+\tfwhm},
\end{equation} 
   
\noindent where $i$ is the inclination of the orbit, \ts \ is secondary eclipse time, \tc \ is the primary transit time, and $\beta$ is the $\delta$ from \citet{Alonso:2018}, but renamed to remove confusion with the transit depth, equal to

\begin{equation}
\label{eq:delta_alonso}
\beta = \sqrt{\frac{1-\left(\frac{\rstar}{\rplanet+\rstar}\right)^2\cos ^2 i}{1-\left(\frac{\rplanet}{\rplanet+\rstar}\right)^2\cos ^2 i}}.
\end{equation} 

Note that the epochs of \ts \ and \tc \ must minimize the numerator in equation \ref{eq:ecosw}, and should be corrected for the observed delay due to the light travel time between the primary and secondary eclipses. 

While these equations are approximate, detailed models like the projected Keplerian orbits used by \exofasttwo \ determine accurate values for $e$ and $\omega_*$ numerically. Then, we can determine \ar \ from equation \ref{eq:arobserved}, and finally \rhostar \ from equation \ref{eq:rearranged_rhostar}. \citet{Eastman:2023} then showed from the definition of $\rhostar \equiv 3\mstar/4\pi\rstar^3$ how its error propagates to other quantities of interest. In particular, they determined the fractional error in \mstar \ is reduced by a factor of 3 as it propagates through \rhostar \ to \rstar.

\begin{equation}
    \label{eq:sigma_rstar_frac}
    \left(\frac{\sigma_{\rstar}}{\rstar}\right)^2 = \left(\frac{\sigma_{\rhostar}}{3\rhostar}\right)^2 + \left(\frac{\sigma_{\mstar}}{3\mstar}\right)^2
\end{equation}

So by determining the mass to 2.3\% (the scatter in the dynamically calibrated \ks-\mstar \ relation) and the stellar density to 2.7\% (a typical constraint achieved with JWST observations of transits and occultations), we can derive the stellar radius to 1\% just from the definition of \rhostar \ and using standard error propagation in equation \ref{eq:sigma_rstar_frac}. 

That is, relying on well-understood physics without any reliance on stellar atmospheric or evolutionary models or a spectroscopic measurement of \teff, our precision is $\sim3\times$ better than IRFM and comparable to the controversial precision claimed by gold standard LBOI observations, or the rare detached double-lined eclipsing binaries. Not only do we precisely determine the properties of specific systems and improve the precision of fundamental planet parameters, but it can also serve as an independent check on LBOI observations and expand the sample of precise stellar radii with which to calibrate future relations similar to \citet{Mann:2015}.

Finally, because we no longer rely on the spectroscopic \teff \ to determine \rstar, \citet{Eastman:2023} also showed that we can leverage this empirically derived radius, a precise distance from \gaia, and the definition of the bolometric flux \fbol,

\begin{equation}
    \label{eq:fbol}
    \fbol \equiv  \sigmasb \teff^4 \left(\frac{\rstar}{d}\right)^2,
\end{equation}

\noindent to determine a precise temperature,

\begin{equation}
    \label{eq:sigma_teff_frac_fbol}
    \left(\frac{\sigma_{\teff}}{\teff}\right)^2 =
    \left(\frac{\sigma_{d}}{2d}\right)^2 + \left(\frac{\sigma_{\fbol}}{4\fbol}\right)^2 + \left(\frac{\sigma_{\rstar}}{2\rstar}\right)^2,
\end{equation}

\noindent where $\sigmasb$ is the Stefan-Boltzmann constant and $d$ is the distance to the star.

While this temperature loosely depends on stellar model atmospheres or bolometric corrections to determine \fbol, \citet{Tayar:2022} showed that a realistic systematic uncertainty in \fbol \ is $\sim$2.4\%. Given that they focused on solar-type stars and M dwarfs are more poorly known, we assume the floor for our low-mass stars is 4\%. Equation \ref{eq:sigma_teff_frac_fbol} shows that the fractional uncertainty in \fbol \ is reduced by a factor of 4 as it propagates to \teff. Thus, even with a 4.0\% systematic floor in \fbol \ and 1\% uncertainty in \rstar, we can determine a 1.25\% uncertainty in \teff --- almost twice as precise as the systematic floors from spectroscopy.

While all these ideas have been done individually before \citep[e.g.,][]{Charbonneau:2005,Huber:2017,Mann:2017,Eastman:2023}, combining them in this way is new and allows many more targets to be measured with a precision that makes them suitable calibration stars. We present a proof of concept for this novel method by analyzing a JWST full-phase light curve of the M dwarf \thisstar \ and its transiting super-Earth GJ 1214 b to determine the stellar radius and temperature to a mere $\sim$1\%.

We discuss the various data used in our analysis, including a JWST full phase light curve, in \S \ref{sec:obs}. Our methods for our models as created by \exofasttwo \ are discussed in \S \ref{sec:methods}. Our results are presented in \S \ref{sec:results}. A discussion of the importance of our research as a whole as well as future applications is given in \S \ref{sec:discussion}.

\section{Observations}
\label{sec:obs}

\subsection{JWST}
\citet{Kempton:2023} observed a full white-light phase curve of \thisstar \ with JWST, which included two secondary eclipses and a primary transit observed with JWST's Mid-Infrared Instrument low resolution spectrometer (MIRI LRS) in the Slitless prism mode with 6.68 s of integration per frame. \thisstar \ was observed continuously from just before the secondary eclipse on 2022 July 20 until just after the next secondary eclipse on 2022 July 22. The JWST data presented in this article were obtained from the Mikulski Archive for Space Telescopes (MAST) at the Space Telescope Science Institute. The specific observations analyzed can be accessed via \dataset[DOI: 10.17909/19mj-5h97]{https://doi.org/10.17909/19mj-5h97}.

While Kempton's white-light phase curve of their observations is available online \citep{Zhang:2023}, we discovered that its time stamps did not match the desired {\tt INT\_MID\_BJD\_TDB} keyword from the data in the archive, resulting in a $\sim$2 minute offset in the observed transit time. \citet{Kempton:2023} were unaware of the {\tt INT\_MID\_BJD\_TDB} keyword, and computed their own \bjdtdb \ using the Barycentric time correction {\tt BARTDELT} keyword. However, due to a known issue in that keyword\footnote{\url{https://jwst-docs.stsci.edu/jwst-calibration-pipeline-caveats/known-issues-with-jwst-data-products}}, their times were incorrect. They noticed the problem late in their analysis and applied a correction to their modeled transit time, but did not update the publicly available light curves (Kempton, et. al., private communication). 

After noticing that an $\sim$8s \ ($\sim$7$\sigma$) difference remained, we ultimately determined the remaining discrepancy to be a long-standing bug in \exofasttwo. The \exofasttwo \ model corrects the observed times to the target's barycentric frame (\tjdtdb) to do its internal modeling (e.g., to account for the light travel time between the primary and secondary eclipse times to constrain the eccentricity), but failed to correct back to the ``observed'' solar system barycenter frame (\bjdtdb), and erroneously reported the times to be in \bjdtdb. \exofasttwo \ now correctly identifies which times are \tjdtdb \ and which are \bjdtdb, but all timestamps of all prior models quoted using \exofasttwo \ are actually \tjdtdb, which would make transit times appear systematically early (by the light travel time) relative to stated \bjdtdb. The results in Table \ref{tab:gj1214} correctly report the transit times. 

Prior to receiving the corrected light curve from \citet{Kempton:2023}, we extracted our own white-light phase curve using the Tiberius pipeline \citep{Kirk:2017, Kirk:2021}. We began by processing the uncalibrated images (\texttt{uncal.fits}) files through stage 1 of the JWST Science Calibration Pipeline (version 1.8.2 \citet{Bushouse:2022}). We followed the default steps for processing MIRI time series observations, apart from the \texttt{jump} step that has been demonstrated to lead to increased noise in time series data \citep[e.g.,][]{Rustamkulov:2023}. Instead, we performed our own outlier detection by comparing all pixels in the time-series to a median-combined integration. Any pixels in the time-series that deviated by $>5\sigma$ from the median pixel value were replaced with the median value. We then proceeded to run our spectral extraction on the outlier cleaned integrations. We began by tracing the centers of the stellar spectra with a quartic polynomial and performed standard aperture photometry with an aperture full width of 8 pixels, equal to the choice of \cite{Kempton:2023}, after removing the background at every row on the detector. The background was measured with a linear polynomial fit to two 6-pixel-wide regions offset by 15 pixels from either side of the extraction aperture. Like \cite{Kempton:2023}, we also saw an exponential ramp at the beginning of our light curve. Following the approach of \cite{Kempton:2023}, we clipped the first 550 integrations (63 minutes) and used a running median with a sliding box of 216 data points to remove outliers. In our study, we decided to remove $>3 \sigma$ outliers by this method. The RMS of the Tiberius-extracted light curve after removing our best-fit transit model, was 302 ppm -- very close to the 306 ppm RMS we achieved for the Kempton-extracted light curve.

\subsection{Other Transit Photometry}
There is a wealth of ground-based observations on \thisstar, beginning with its discovery \citep{Charbonneau:2009}. Despite minor systematics, we opted to include their Fred Lawrence Whipple Observatory (FLWO) data and their MEarth data as the 14 yr baseline led to an improvement in the precision of the period by a factor of 10,000, relative to the JWST light curve alone. However, their overall contribution to the total S/N of the combined light curve was dwarfed by JWST; thus the minor systematics have a negligible effect on the determination of \rhostar. 

Very Large Telescope (VLT) data from \citet{Berta:2011} was nominally very precise, but showed strong systematic errors by eye. We modeled the VLT data including quadratic detrending as done in \citet{Berta:2011}. However, their inclusion did not measurably impact either the values or precision of any parameter. Because of that, and out of a concern for the significant increase in run time for our models, we did not include them in our final analysis.

We also tried including Spitzer light curves from \citet{Cloutier:2021}, who reextracted observations by \citet{Fraine:2013} and \citet{Gillon:2017}. The Spitzer data contained large systematics that needed to be removed with heroic effort. \cite{Gillon:2014} detrended each light curve with an optimized set of parameters specific to each day that data was collected, whereas \citet{Cloutier:2021} used a more sophisticated approach that relied on Gaussian processes for detrending. We found that the addition of the detrended data from Cloutier (2023, private communication) did not significantly improve the precision except for the period, but our global model showed a systematic shift of $\sim$0.7$\sigma$ in \rhostar, \rstar, and \teff. Out of a concern that the systematic removal process may have been imperfect, we chose to discard the Spitzer data in our global analysis. 

Finally, we checked for \tess \ data, but according to \tess-point, \thisstar \ was not observed in any sector to date, nor will it be included in any planned sector through sector 83 ending on 2024 October 1.

\subsection{Stellar Photometry}
\begin{figure}[!htbp]
  \begin{center}
    \includegraphics[width=3.5in, trim=2.5cm 0cm 0cm 0cm, clip]{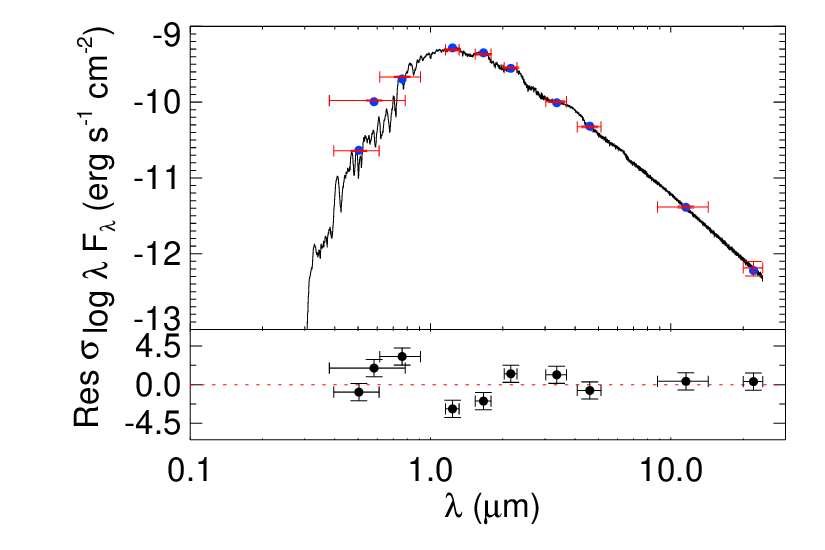} 
    \caption{The upper panel shows the SED model for \thisstar \ of brightness versus wavelength in microns. The black line represents the  model's atmosphere and the red horizontal bars represent wavelength span of the observations in the star. The blue points represent the model and the average over the span of the width of the red horizontal bars. The red vertical bars represent the uncertainty in the measurement; the bottom panel represents the model's residuals of observed minus calculated divided by the uncertainty.
}
      \label{fig:sed}

  \end{center}
\end{figure}

The broadband photometry used for the SED models is summarized in Table \ref{tab:LitProps} and shown in Figure \ref{fig:sed}. The uncertainties were rounded based on standard procedures where we assume a 0.02-0.03 magnitude systematic floor in determining the absolute magnitude. We used data from \gaia \ Data Release 3 (DR3), Two Micron All Sky Survey (2MASS), and Wide Field Infrared Survey Explorer (WISE). Note that \thisstar \ was not automatically cross matched in TICv8.2 \citep{Stassun:2018}, so WISE and \gaia \ DR3 matching was done by determining the star with the smallest Ks-WISE1 color within 30'', as is now automatically codified in the \exofasttwo \ utility \texttt{MKTICSED}. The corresponding IDs are included in Table \ref{tab:LitProps}.

\begin{table}
\scriptsize
\setlength{\tabcolsep}{2pt}
\centering
\caption{Literature and Measured Properties for \thisstar}
\begin{tabular}{llcc}
  \hline
  \hline
Other identifiers\dotfill & \\
\multicolumn{3}{c}{GJ 1214} \\
\multicolumn{3}{c}{LHS 3275, TIC 467929202}\\
\multicolumn{3}{c}{2MASS J17151894+0457496, \gaia \ DR2 4393265392167891712} \\
\multicolumn{3}{c}{\gaia \ DR3 4393265392168829056, ALLWISE J171519.33+045742.2}\\
\hline
\hline
Parameter & Description & Value & Source\\
\hline 
$\alpha_{J2000}^\ddagger$\dotfill	&Right Ascension (RA)\dotfill & 17:15:18.934& 1	\\
$\delta_{J2000}^\ddagger$\dotfill	&Declination (Dec)\dotfill & 04:57:50.067& 1	\\
\\
$l$\dotfill     & Galactic Longitude\dotfill & 26.16210287$^\circ$ & 1\\
$b$\dotfill     & Galactic Latitude\dotfill & 23.60692189$^\circ$ & 1\\
\\
${\rm G}$\dotfill     & \gaia \ $G$ mag.\dotfill & 12.997$ \pm0.020$ & 1 \\
G$_{\rm BP}$\dotfill & \gaia \ BP mag.\dotfill & 14.912$ \pm0.020$ & 1 \\
G$_{\rm RP}$\dotfill     & \gaia \ RP mag.\dotfill & 11.683$ \pm0.020$ & 1 \\
\\
J\dotfill			& 2MASS J mag.\dotfill & 9.75 $\pm$ 0.024	& 2	\\
H\dotfill			& 2MASS H mag.\dotfill & 9.094 $\pm$ 0.024	    &  2	\\
K$_S$\dotfill		& 2MASS ${\rm K_S}$ mag.\dotfill & 8.782 $\pm$ 0.020&  2	\\
\textit{WISE1}\dotfill	& \textit{WISE1} mag.\dotfill & 8.598 $\pm$ 0.030  & 3 \\
\textit{WISE2}\dotfill	& \textit{WISE2} mag.\dotfill & 8.442 $\pm$ 0.030  & 3 \\
\textit{WISE3}\dotfill	& \textit{WISE3} mag.\dotfill & 8.254 $\pm$ 0.030  & 3 \\
\textit{WISE4}\dotfill	& \textit{WISE4} mag.\dotfill & 8.084 $\pm$ 0.236  & 3 \\
\\
$\mu_{\alpha}$\dotfill		& \gaia \ DR2 proper motion\dotfill & 580.202 $\pm$ 0.059 & 1 \\
                            & \hspace{3pt} in RA (mas yr$^{-1}$)	&  \\
$\mu_{\delta}$\dotfill		& \gaia \ DR2 proper motion\dotfill 	&  -749.713 $\pm$ 0.047 &  1 \\
                            & \hspace{3pt} in DEC (mas yr$^{-1}$) &  \\
$\pi^\parallel$\dotfill & \gaia \ Parallax (mas) \dotfill & 68.299 $\pm$  0.065 &  1 \\
$RV$\dotfill & Systemic radial \hspace{9pt}\dotfill  & $20.91\pm0.66$  & 1 \\
     & \hspace{3pt} velocity (\kmps)  & \\
$d$\dotfill & Distance (pc)\dotfill & 14.642$\pm0.014$ & 1 \\
\hline
\end{tabular}
\begin{flushleft}
 \footnotesize{ \textbf{\textsc{NOTES:}}
 The uncertainties of the photometry have a systematic error floor applied. \\
 $\ddagger$ RA and Dec are in epoch J2000. The coordinates come from Vizier where the \gaia \ RA and Dec have been precessed and corrected to J2000 from epoch J2015.5.\\
 $\parallel$ Values from \gaia \ DR3 are raw because they could not be corrected for the systematic offset as prescribed by \citet{Lindegren:2021}. There was no value for {\tt nu\_eff\_used\_in\_astrometry} in DR3 for this star. \\
 
 References are: $^1$\citet{Gaia:2020},$^2$\citet{Cutri:2003}, $^3$\citet{Cutri:2012}\\ 
}
\end{flushleft}
\label{tab:LitProps}
\end{table}

\subsection{Radial Velocities}

We used the same 165 spectra from the High Accuracy Radial velocity Planet Searcher \citep[HARPS;][]{Mayor:2003} spectra from \citet{Cloutier:2021}, which span from 2009 July 24 to 2019 Sepetember 2 and include a reanalysis of the discovery Radial Velocities from \citet{Charbonneau:2009}, and their own observations. In 2015 June, the HARPS fiber was upgraded. We modeled the preupgraded and postupgraded data with separate zero-points and jitter terms. 

\section{Methods}
\label{sec:methods}
We performed a global model of the \thisstar \ system using \exofasttwo \ and the data described in \S \ref{sec:obs}. We used the default options as outlined in \citet{Eastman:2019}, except the following. 

We opted not to use the default MIST evolutionary model and instead used a prior on the $k_S$ apparent magnitude and the recently integrated \citet{Mann:2019} \ks-\mstar \ relation, as the dynamical relation is much more reliable for low-mass stars. We fit a spline \citep{Vanderburg:2014} and changed the default knot spacing from 0.75 days to 0.4 days to remove the phase-curve variability of the JWST extracted light curve. The value of the knot spacing was an appropriate compromise between being much smaller than the period ($\sim0.25\times$) in order to take out the phase variations, but much larger than the transit duration ($\sim10\times$) so as to not bias the transit shape. 

We fit a secondary eclipse to the JWST data due the fact that it was noticeably observed in the light curve and provided a strong constraint on the eccentricity (see equations \ref{eq:ecosw} to \ref{eq:delta_alonso}). We disabled the constraint from the limb-darkening tables \citep{Claret:2017} because the light curve was capable of constraining this on its own, there is no constraint given by \citet{Claret:2017} for the JWST MIRI LRS Slitless prism, and theoretically predicted limb-darkening coefficients can be especially poor for low-mass stars \citep{Patel:2022}. 

We detrended FLWO and MEarth light curves with airmass simultaneously with the fit, following the procedure in the discovery paper \citep{Charbonneau:2009}.

We used a parallax prior from \gaia \ DR3 \citep{Gaia:2020}. Note that we did not correct for the systematic offset as prescribed by \citet{Lindegren:2021} because there was no value for the required effective wavenumber of the source used in the astrometric solution \texttt{nu\_eff\_used\_in\_astrometry} for \thisstar \ in DR3, though we do not expect this to be a significant source of error \citep{Stassun:2021}. Further, \citet{El-Badry:2021} showed that the \gaia \ DR3 uncertainties can be underestimated by as much as 30\% for bright ($G \lesssim 18$) stars, which we accounted for by inflating the error in the parallax by 30\%. However, because the uncertainty is so small (\gjdistanceerr\%), the impact on the rest of the fit is negligible. Our inflated parallax differs by 0.5$\sigma$ from the Early Data Release 3 (EDR3) value used by \citet{Cloutier:2021}. We also used a metalicity prior from \citet{Cloutier:2021} and an upper limit on the extinction from \citet{Schlafly:2011} and \citet{Schlegel:1998}. These priors are summarized in Table \ref{tab:Priors}.

Measuring bolometric fluxes is difficult. This is especially true for low-mass stars which have poorly understood atmospheres. Our method of computing bolometric flux relies on the SED model, which requires that we understand the filter profiles and zero-points for each of the modeled bandpasses. We chose to use the MIST bolometric correction tables, which use detailed filter profiles, including the latest updated profiles for \gaia \ DR3.

However, remaining systematics in filter profiles, zero-points, and atmospheric models must be accounted for. In order to do this, we used three independent procedures. First, we inflated the per-point measurement uncertainites to a floor of between 0.02 to 0.03 magnitudes to ensure the proper relative weighting of each magnitude. Next, \exofasttwo \ fits a term called $\sigma_{\rm SED}$ (as seen in Table \ref{tab:gj1214}) which scales all SED photometry errors together so that the uncertainties in the data are well matched to the model ($\chi^2_R\sim1$).

Finally, \exofasttwo \ includes a systematic error floor in the SED model to prevent over-constraining \fbol \ and \teff \ from the SED. It is implemented by fitting $F_{\rm Bol,SED}$ and $T_{eff,SED}$ -- constrained only by the SED -- and the same quantities again, $F_{Bol}$ and \teff, but used everywhere else in the fit. They are then linked to each other with penalties that account for our presumed systematic error floors, $\sigma_{T_{\rm eff,sys}}$, and $\sigma_{F_{\rm Bol,sys}}$,

\begin{equation}
\label{eq:teffsed}
    \ln{\mathscr{L}} += -0.5\left(\frac{T_{\rm eff, SED}-T_{\rm eff}}{\sigma_{T_{\rm eff,sys}}}\right)^2,
\end{equation}

\noindent and

\begin{equation}
\label{eq:fbollike}
    \ln{\mathscr{L}} += -0.5\left(\frac{F_{\rm Bol, SED} - F_{\rm Bol}}{\sigma_{F_{\rm Bol,sys}}}\right)^2.
\end{equation}

By default, \exofasttwo \ assumes $\sigma_{T_{\rm eff,sys}}=2\%$ and $\sigma_{F_{\rm Bol,sys}}$=2.4\%, as recommended by \citet{Tayar:2022}. However, we further inflated $\sigma_{F_{\rm Bol,sys}}$ to 4\% to account for larger expected systematics in M dwarfs. This gives us our final precision in \fbol \ of \gjfbolerr\%.

Note that since the density of the star coupled with the mass from \citet{Mann:2019} is the primary constraint on the radius of the star, the impact of the SED model on \rstar \ is negligible. This means that systematic errors imparted by the SED will only impact our measurement of \teff. We also note that the final uncertainty in \teff \ can be better than $\sigma_{T_{\rm eff,sys}}$ when it is determined from $\sigma_{d}$, $\sigma_{\fbol}$, and $\sigma_{\rstar}$ through equation \ref{eq:sigma_teff_frac_fbol}, as it is here and is typically the case any time $\sigma_{\rstar}$ is small.

\startlongtable
\begin{deluxetable}{lcccc}
\tablecaption{Priors in the \texttt{EXOFASTv2} Global Fits for GJ 1214\label{tab:Priors}}
\tablehead{\colhead{Param.} & \colhead{Description} & \colhead{Prior} & \colhead{Source}}
\startdata
$\varpi$ & Parallax [mas] & $\mathcal{N}$(68.299, 0.0845) & 1 \\
$\left[\mathrm{Fe/H}\right]$ & Metallicity [dex] & $\mathcal{N}$(0.29 0.12) & 2\\ 
$k_s$ & Apparent $K_s$ [mag]  & $\mathcal{N}$(8.782, 0.020)  & 3 \\ 
$A_V$ & Reddening [mag] & $\mathcal{U}$[0, 0.4668507] & 4,5\\
\enddata
\tablenotetext{}{Here, $\mathcal{N}$(a,b) is a Gaussian distribution centered at a with a standard deviation of b, and $\mathcal{U}$[c, d] is a uniform distribution bounded inclusively between c and d. \\References are:$^1$\citet{Gaia:2020},$^2$\citet{Cloutier:2021}, $^3$\citet{Cutri:2003}, $^4$\citet{Schlafly:2011}, $^5$\citet{Schlegel:1998}}
\end{deluxetable}

\section{Results}

\label{sec:results}
\begin{figure}[!htbp]
  \begin{center}
    \includegraphics[width=3.5in, trim=0cm 0cm 0cm 20cm, clip]{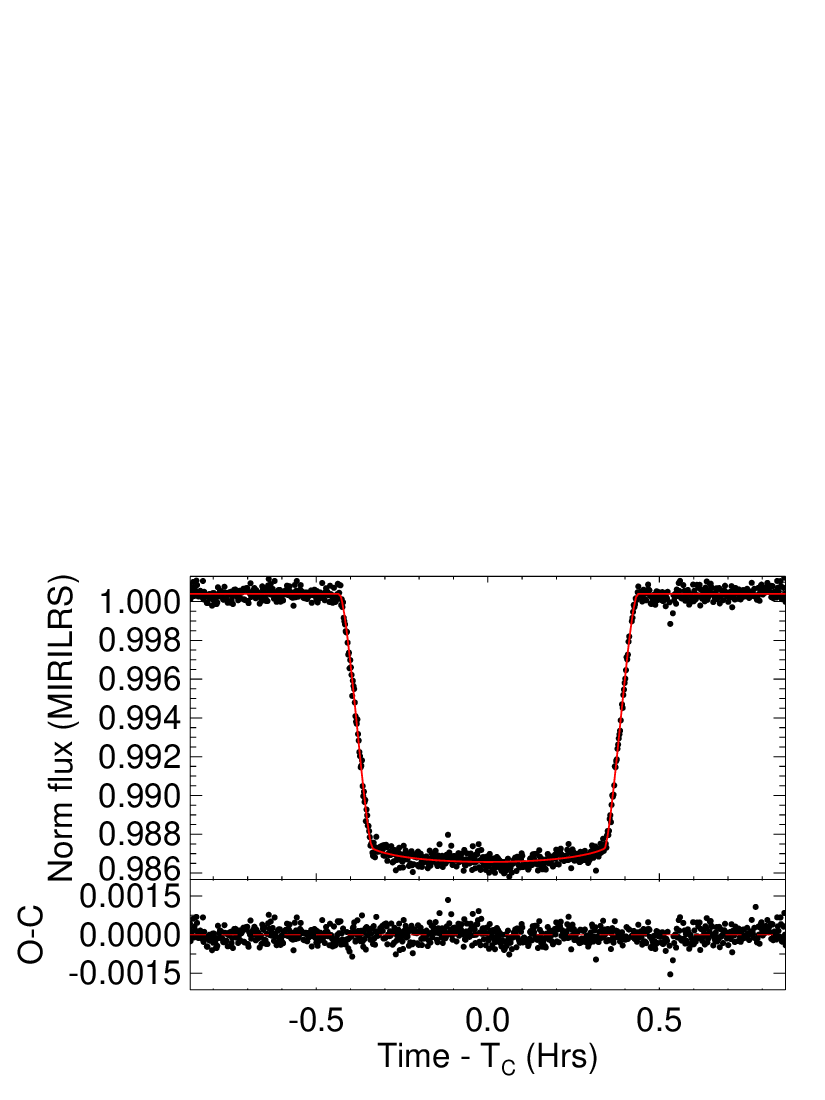}
    \caption{
    The upper panel shows the primary JWST transit of \thisstar b. The black points represent each respective data point, the red line represents the best-fit model using \exofasttwo, and the lower panel representing the model's residuals of observed minus calculated. The JWST light curve is available as data behind the figure in the online Journal.
    }
    \label{fig:transit}
  \end{center}
\end{figure}

\begin{figure}[!htbp]
  \begin{center}
    \includegraphics[width=3.5in,trim=0.5cm 0cm 0cm 19.5cm, clip]{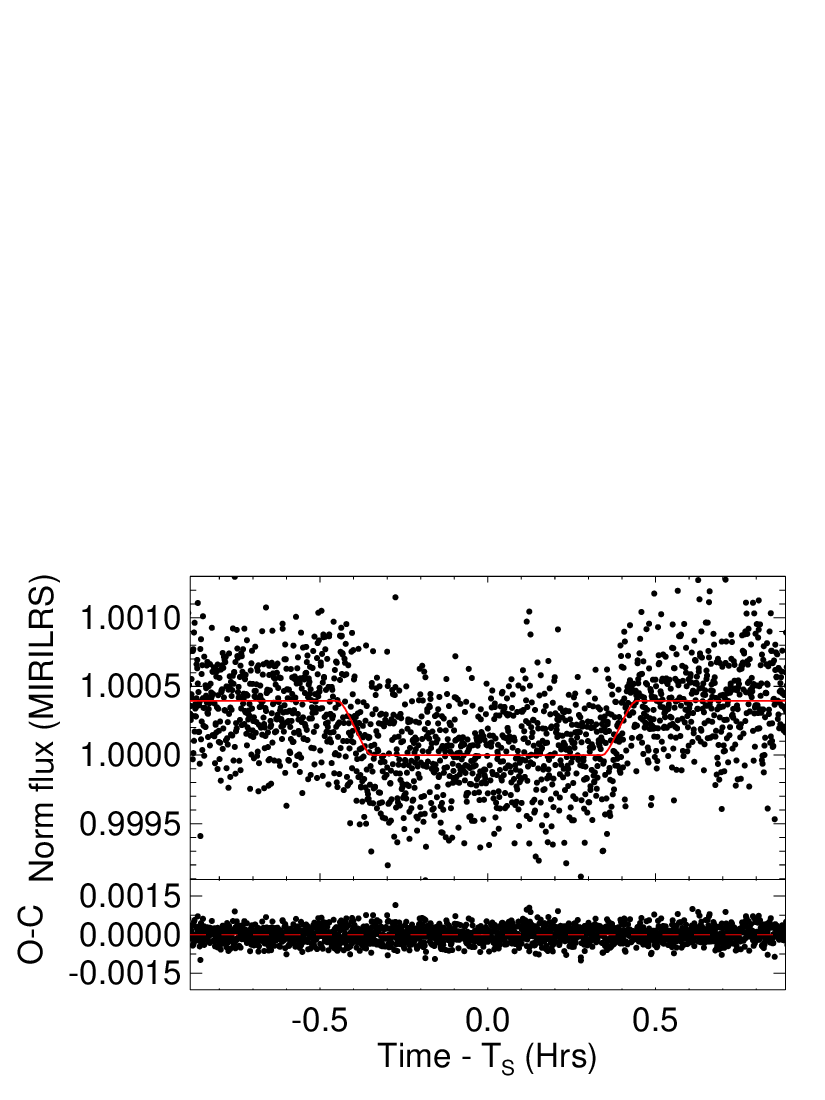} 
    \caption{
    Same as Figure \ref{fig:transit}, but for the JWST secondary eclipse. Note the transit has been normalized so that the stellar flux (i.e., during the secondary eclipse) is 1.
    }
    \label{fig:secondary}
  \end{center}
\end{figure}

\begin{figure}[!htbp]
  \begin{center}
    \includegraphics[width=3.5in,trim=2cm 0cm 1cm 0cm, clip]{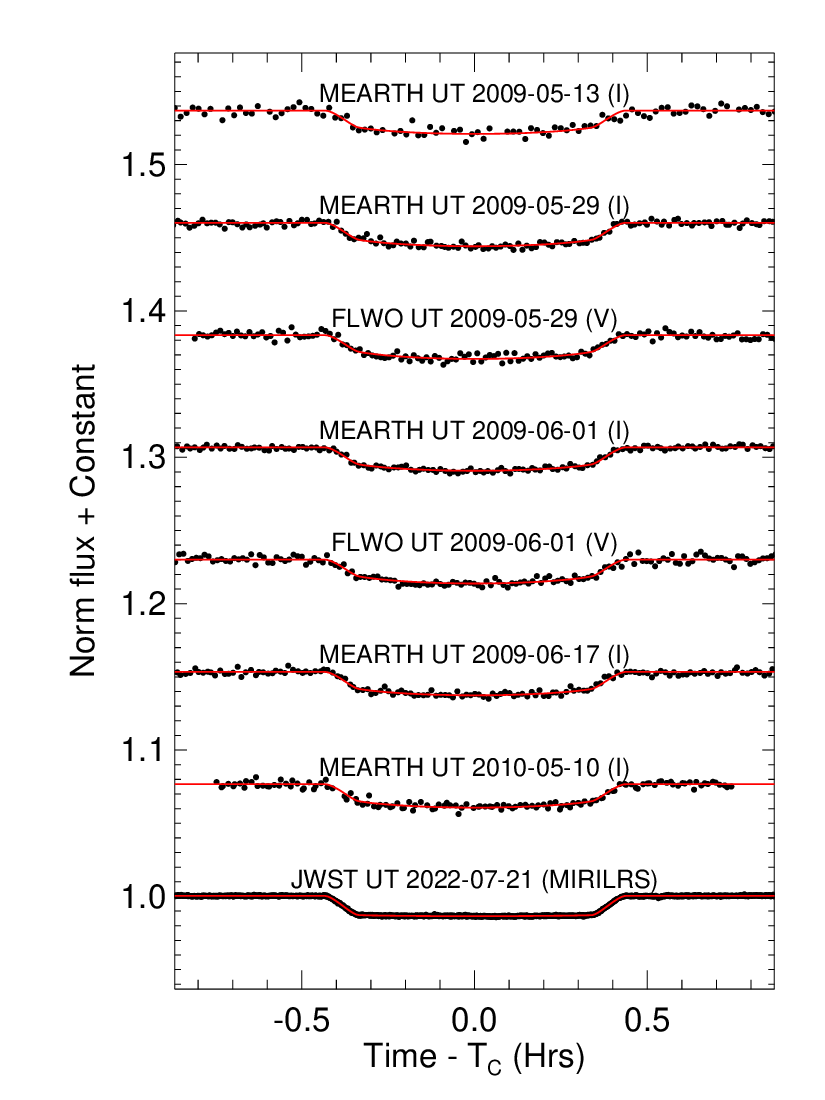} 
    \caption{
    All primary transit light curves used in our global model, plotted with an arbitrary offset. The black points represent each respective data point and the red lines represent the best-fit model using \exofasttwo. Labels above each transit state the telescope, epoch, and filter used for the observation.
    }
    \label{fig:stacked_transit}
  \end{center}
\end{figure}

All fitted and derived parameters from our analysis of \thisstar \ are summarized in Table \ref{tab:gj1214}. Figures \ref{fig:transit} and \ref{fig:secondary} show the primary transit and secondary eclipse from JWST, respectively. Figure \ref{fig:stacked_transit} shows all light curves used for our analysis (i.e., JWST, MEarth, and FLWO). Notably, our analysis dramatically improves the precision for \rstar \ (\gjrstarfac$\times$), \teff \ (\gjtefffac$\times$), and \rhostar \ (\gjrhostarfac$\times$) compared to the next best result by \citet{Cloutier:2021}, as shown in the ``Improvement'' column in Table \ref{tab:gj1214}.

The secondary transit timing also constrains the eccentricity of the orbit \gjeupfac \ times better, despite \citet{Cloutier:2021} using over a decade of HARPS data. We note that the eccentricity and its uncertainty from our global model ($e=$\gje) are in excellent agreement with values numerically propagated using equations \ref{eq:ecosw}--\ref{eq:delta_alonso}, and our MCMC chain values for inputs ($e=0.0058_{-0.0040}^{+0.0073}$).

Finally, the improved stellar parameters propagate to improved accuracy and precision in the planetary parameters, including the planet radius \rplanet \ (\gjrpfac$\times$) and the incident flux on the planet, \fave \ (\gjinsolfac$\times$), which is critical for understanding exoplanets and their climates. 

We note that the tidal circularization timescale reported by \exofasttwo \ by default uses $Q=10^6$ computed from \citet[][eq. 3]{Adams:2006}, but a more appropriate value for a rocky planet like \thisstar b is $Q=100$. We have manually corrected the tidal circularization timescale to \gjtcirc \ Myr in Table \ref{tab:gj1214}, assuming $Q=100$. Given that, we expect the system to be tidally circularized, consistent with our eccentricity of \gje. 

All of our values are consistent with \citet{Cloutier:2021} within $3\sigma$, except the depth (\gjdeltasigdiff$\sigma$), which is expected due to differences in observed wavelength and stellar variability. We note that the discrepancy is negligible when propagated to the planetary radius. A few other values differ by more than $1\sigma$. \teff \ is \gjteffsigdiff$\sigma$ lower than their spectroscopically determined value. \teff-derived quantities (\teq, \fave) are similarly discrepant. This is a relatively small difference given that the two methods are nearly completely independent determinations of the stellar temperature, though we expect ours is more direct and therefore more accurate. Our inclination and period also differ by \gjisigdiff$\sigma$ and \gjperiodsigdiff$\sigma$, respectively.

We note that the limb darkening for our MIRI LRS white-light curve ($u_1=$ \gjuoneMIRI, $u_2=$ \gjutwoMIRI) is better constrained here than from theoretical predictions \citep{Patel:2022}. Because of the strong covariance between limb darkening and \ar \ (and ultimately \rhostar), it is important that we have accurate values with uncertainties accounted for within the global model.

\startlongtable
\begin{deluxetable*}{lcccc}
\tablecaption{Median values and 68\% confidence interval for \thisstar.}
\tablehead{\colhead{~~~Parameter} & \colhead{Description} & \colhead{This Work} & \colhead{\citet{Cloutier:2021}} & \colhead{Improvement$^1$}}
\startdata
\smallskip\\\multicolumn{2}{l}{Stellar Parameters:}&\smallskip\\
~~~~$M_*$\dotfill &Mass (\msun)\dotfill                                        & \gjmstar                 & $0.178 \pm 0.010$             & \gjmstarfac    \\
~~~~$R_*$\dotfill &Radius (\rsun)\dotfill                                      & \gjrstar                 & $0.215 \pm 0.008$             & \gjrstarfac    \\
~~~~$R_{*,SED}$\dotfill &Radius$^{2}$ (\rsun)\dotfill                          & \gjrstarsed              & --                            & --             \\
~~~~$L_*$\dotfill &Luminosity (\lsun)\dotfill                                  & \gjlstar                 & $0.0046^{+0.0007}_{-0.0006}$  & \gjlstarfac    \\
~~~~$F_{Bol}$\dotfill &Bolometric Flux (cgs)\dotfill                           & \gjfbol                  & --                            & --             \\
~~~~$\rho_*$\dotfill &Density (cgs)\dotfill                                    & \gjrhostar               & $25.4^{+3.5}_{-3.0}$          & \gjrhostarfac  \\
~~~~$\log{g}$\dotfill &Surface gravity (cgs)\dotfill                           & \gjlogg                  & $5.026 \pm 0.040$             & \gjloggfac     \\
~~~~$T_{\rm eff}$\dotfill &Effective Temperature (K)\dotfill                   & \gjteff                  & $3250 \pm 100$                & \gjtefffac     \\ 
~~~~$T_{\rm eff,SED}$\dotfill &Effective Temperature$^{2}$ (K)\dotfill         & \gjteffsed               & --                            & --             \\ 
~~~~$[{\rm Fe/H}]$\dotfill &Metallicity (dex)\dotfill                          & \gjfeh                   & $0.29 \pm 0.12$               & \gjfehfac      \\
~~~~$K_s$\dotfill &Absolute Ks (mag) \dotfill                                  & \gjksabs                 & $7.956 \pm 0.020$             & \gjksabsfac    \\
~~~~$k_s$\dotfill &Apparent Ks (mag) \dotfill                                  & \gjksapp                 & --                            & --             \\
~~~~$A_V$\dotfill &V-band extinction (mag)\dotfill                             & \gjav                    & --                            & --             \\
~~~~$\sigma_{SED}$\dotfill &SED photometry error scaling \dotfill              & \gjsigmased              & --                            & --             \\
~~~~$\varpi$\dotfill &Parallax (mas)$^{3}$\dotfill                             & \gjparallax              & $68.348 \pm 0.056$            & \gjparallaxfac \\
~~~~$d$\dotfill &Distance (pc)\dotfill                                         & \gjdistance              & $14.631 \pm 0.012$            & \gjdistancefac \\
\smallskip\\\multicolumn{2}{l}{Planetary Parameters:}&\smallskip\\
~~~~$P$\dotfill &Period (days)\dotfill                                         & \gjperiod                & $1.58040433 \pm 0.00000013$   & \gjperiodfac   \\
~~~~$R_P$\dotfill &Radius (\re)\dotfill                                        & \gjrp                    & $2.733^{+0.050}_{-0.052}$     & \gjrpfac       \\
~~~~$M_P$\dotfill &Mass (\mearth)\dotfill                                      & \gjmp                    & $8.17 \pm 0.43$               & \gjmpfac       \\
~~~~$T_C$\dotfill &Time of conjunction$^{4}$ (\bjdtdb)\dotfill                 & \gjtc                    & $2459639.78092 \pm 0.00033$ & \gjtcfac         \\
~~~~$T_T$\dotfill &Time of min proj sep$^{5}$ (\bjdtdb)\dotfill                & \gjtt                    & --                            & --             \\
~~~~$a$\dotfill &Semi-major axis (AU)\dotfill                                  & \gja                     & $0.01490 \pm 0.00026$         & \gjafac        \\
~~~~$i$\dotfill &Inclination (Degrees)\dotfill                                 & \gji                     & $88.7 \pm 0.1$                & \gjifac        \\
~~~~$e$\dotfill &Eccentricity \dotfill                                         & \gje                     & --                            & --             \\
~~~~$e$\dotfill &Eccentricity 95\% Upper Limit \dotfill                        & \gjeup                   & $<0.063$                      & \gjeupfac      \\
~~~~$\omega_*$\dotfill &Argument of Periastron (Degrees)\dotfill               & \gjomega                 & --                            & --             \\
~~~~$T_{\rm eq}$\dotfill &Equilibrium temperature$^{6}$ (K)\dotfill            & \gjteq                   & $596 \pm 19$                  & \gjteqfac      \\
~~~~$\tau_{\rm circ}$\dotfill &Tidal circ timescale (Myr)\dotfill              & \gjtcirc                 & $<1$                      & --             \\
~~~~$K$\dotfill &RV semi-amplitude (m/s)\dotfill                               & \gjk                     & $14.36 \pm0.53$               & \gjkfac        \\
~~~~$R_P/R_*$\dotfill &Radius of planet in stellar radii \dotfill              & \gjp                     & $0.11677 \pm 0.00017$         & \gjpfac        \\
~~~~$a/R_*$\dotfill &Semi-major axis in stellar radii \dotfill                 & \gjar                    & $14.85 \pm 0.16$              & \gjarfac       \\
~~~~$\delta$\dotfill &$\left(R_P/R_*\right)^2$ $^{3}$ \dotfill                 & \gjdelta                 & $0.013635 \pm 0.000040$       & \gjdeltafac    \\
~~~~$\delta_{\rm I}$\dotfill &Transit depth in I (fraction)\dotfill            & \gjdeltaI                & --                            & --             \\
~~~~$\delta_{\rm MIRI}$\dotfill &Transit depth in MIRI LRS (fraction)\dotfill  & \gjdeltaMIRI             & --                            & --             \\
~~~~$\delta_{\rm V}$\dotfill &Transit depth in V (fraction)\dotfill            & \gjdeltaV                & --                            & --             \\
~~~~$\tau$\dotfill &Ingress/egress transit duration (days)\dotfill             & \gjtau                   & --                            & --             \\
~~~~$T_{14}$\dotfill &Total transit duration (days)\dotfill                    & \gjtonefour              & --                            & --             \\
~~~~$T_{FWHM}$\dotfill &FWHM transit duration (days)\dotfill                   & \gjtfwhm                 & --                            & --             \\
~~~~$b$\dotfill &Transit impact parameter \dotfill                             & \gjb                     & $0.325 \pm 0.025$             & \gjbfac        \\
~~~~$b_S$\dotfill &Eclipse impact parameter \dotfill                           & \gjbs                    & --                            & --             \\
~~~~$\tau_S$\dotfill &Ingress/egress eclipse duration (days)\dotfill           & \gjtaus                  & --                            & --             \\
~~~~$T_{S,14}$\dotfill &Total eclipse duration (days)\dotfill                  & \gjtonefours             & --                            & --             \\
~~~~$T_{S,FWHM}$\dotfill &FWHM eclipse duration (days)\dotfill                 & \gjtfwhms                & --                            & --             \\
~~~~$\delta_{S,2.5\mu m}$\dotfill &BB eclipse depth at 2.5$\mu$m (ppm)\dotfill & \gjdeltatwofive          & --                            & --             \\ 
~~~~$\delta_{S,5.0\mu m}$\dotfill &BB eclipse depth at 5.0$\mu$m (ppm)\dotfill & \gjdeltafivezero         & --                            & --             \\ 
~~~~$\delta_{S,7.5\mu m}$\dotfill &BB eclipse depth at 7.5$\mu$m (ppm)\dotfill & \gjdeltasevenfive        & --                            & --             \\ 
~~~~$\rho_P$\dotfill &Density (cgs)\dotfill                                    & \gjrhop                  & $2.20^{+0.17}_{-0.16}$        & \gjrhopfac     \\
~~~~$logg_P$\dotfill &Surface gravity (cgs) $^{3}$\dotfill                     & \gjloggp                 & $3.026 \pm 0.028$             & \gjloggpfac    \\
~~~~$\Theta$\dotfill &Safronov Number \dotfill                                 & \gjsafronov              & --                            & --             \\
~~~~$\fave$\dotfill &Incident Flux (\fluxcgs) $^{3}$\dotfill                   & \gjinsol                 & $0.0286 \pm 0.0035$           & \gjinsolfac    \\
~~~~$T_P$\dotfill &Time of Periastron (\tjdtdb)\dotfill                        & \gjtp                    & --                            & --             \\
~~~~$T_S$\dotfill &Time of eclipse (\bjdtdb)\dotfill                           & \gjts                    & --                            & --             \\
~~~~$T_E$\dotfill &Time of sec min proj sep (\bjdtdb)\dotfill                  & \gjte                    & --                            & --             \\
~~~~$T_A$\dotfill &Time of Ascending Node (\tjdtdb)\dotfill                    & \gjta                    & --                            & --             \\
~~~~$T_D$\dotfill &Time of Descending Node (\tjdtdb)\dotfill                   & \gjtd                    & --                            & --             \\
~~~~$V_c/V_e$\dotfill & Scaled velocity $^{7}$ \dotfill                        & \gjvcve                  & --                            & --             \\
~~~~$e\cos{\omega_*}$\dotfill & \dotfill                                       & \gjecosw                 & --                            & --             \\
~~~~$e\sin{\omega_*}$\dotfill & \dotfill                                       & \gjesinw                 & --                            & --             \\
~~~~$M_P\sin i$\dotfill &Minimum mass (\mj) $^{3}$\dotfill                     & \gjmsini                 & $8.17 \pm 0.43$               & \gjmsinifac    \\
~~~~$M_P/M_*$\dotfill &Mass ratio $^{3}$ \dotfill                              & \gjq                     & $0.000138 \pm 0.000011$       & \gjqfac        \\
~~~~$d/R_*$\dotfill &Separation at mid transit \dotfill                        & \gjdr                    & --                            & --             \\
~~~~$P_T$\dotfill &A priori non-grazing transit prob \dotfill                  & \gjpt                    & --                            & --             \\
~~~~$P_{T,G}$\dotfill &A priori transit prob \dotfill                          & \gjptg                   & --                            & --             \\
~~~~$P_S$\dotfill &A priori non-grazing eclipse prob \dotfill                  & \gjps                    & --                            & --             \\
~~~~$P_{S,G}$\dotfill &A priori eclipse prob \dotfill                          & \gjpsg                   & --                            & --             \\
\smallskip\\\multicolumn{2}{l}{Wavelength Parameters:}\smallskip\\
~~~~$u_{1,I}$\dotfill &Linear limb-darkening coeff \dotfill                    & \gjuoneI                 & --                            & --             \\
~~~~$u_{2,I}$\dotfill &Quadratic limb-darkening coeff \dotfill                 & \gjutwoI                 & --                            & --             \\
~~~~$u_{1,V}$\dotfill &Linear limb-darkening coeff \dotfill                    & \gjuoneV                 & --                            & --             \\
~~~~$u_{2,V}$\dotfill &Quadratic limb-darkening coeff \dotfill                 & \gjutwoV                 & --                            & --             \\
~~~~$u_{1,MIRI}$\dotfill &Linear limb-darkening coeff \dotfill                 & \gjuoneMIRI              & --                            & --             \\
~~~~$u_{2,MIRI}$\dotfill &Quadratic limb-darkening coeff \dotfill              & \gjutwoMIRI              & --                            & --             \\
~~~~$\delta_{S}$\dotfill &Measured eclipse depth (ppm)\dotfill                 & \gjdeltas                & --                            & --             \\
\smallskip\\\multicolumn{2}{l}{HARPS Parameters:}\smallskip\\
~~~~$\gamma_{\rm rel}$\dotfill &Pre-upgrade Relative RV Offset (m/s)\dotfill   & \gjgammapre              & $-3.09^{+1.65}_{-1.66}$       & \gjgammaprefac \\
~~~~$\sigma_J$\dotfill &Pre-upgrade RV Jitter (m/s) $^{3}$\dotfill             & \gjsigmajpre             & $3.52 \pm 0.56$               & \gjsigmajprefac\\
~~~~$\sigma_J^2$\dotfill &Pre-upgrade RV Jitter Variance $^{3}$\dotfill        & \gjjittervarpre          & $12.4 \pm 4.1$                & \gjjittervarprefac\\
~~~~$\gamma_{\rm rel}$\dotfill &Post-upgrade Relative RV Offset (m/s) \dotfill & \gjgammapost             & $-0.61^{+1.76}_{-1.90}$       & \gjgammapostfac \\
~~~~$\sigma_J$\dotfill &Post-upgrade RV Jitter (m/s) $^{3}$\dotfill            & \gjsigmajpost            & $2.25 \pm0.71$                & \gjsigmajpostfac \\
~~~~$\sigma_J^2$\dotfill &Post-upgrade RV Jitter Variance $^{3}$ \dotfill      & \gjjittervarpost         & $5.1 \pm 3.9$                 & \gjjittervarpostfac\\
\label{tab:gj1214}
\enddata
\tablenotetext{}{See Table 3 in \citet{Eastman:2019} for a detailed description of all parameters}
\tablenotetext{1}{The improvement is calculated as the average error bar from Cloutier et al. (2021)
divided by the average error bar from our analysis. Values less than one are less precise, largely
due to more pessimistic priors.}
\tablenotetext{2}{This value ignores the systematic error and is for reference only}
\tablenotetext{3}{\citet{Cloutier:2021} values were not explicitly stated in their paper but we derived them from their other parameters assuming uncorrelated error bars with a Gaussian width of their average error bar}
\tablenotetext{4}{Time of conjunction is commonly reported as the "transit time"}
\tablenotetext{5}{Time of minimum projected separation is a more correct "transit time"}
\tablenotetext{6}{Assumes no albedo and perfect redistribution}
\tablenotetext{7}{See \citet{Eastman:2024} for more information regarding this quantity}

\end{deluxetable*}

\section{Systematics}
\label{sec:systematics}

The residuals in Figure \ref{fig:transit} show clear signs of correlated noise that could be due to star spots or other systematics. To assess their impact on the transit duration, we shifted the residuals from our final fit by -30, -20, -10, +10, +20, and +30 minutes, added back our best-fit model, and refit it, following exactly the same procedure as our original fit. Among these six fits, the largest discrepancy in \rstar, \teff, or \rhostar \ was $\sim$1$ \sigma$. If our fits were completely independent measurements of the same quantities, such a discrepancy is statistically likely, suggesting that the correlated noise has no statistically significant impact on our inferred values. We also ran a fit that simultaneously detrended against x position, y position, and time, but the impact on the final results was negligible ($<0.1$$\sigma$).

\citet{Moran:2023} noted $\sim$6$\sigma$ discrepancies in \ar \ for JWST observations of a different star, GJ 486. Their comparison only changed the extraction pipeline, suggesting that light curve extraction systematics may be significant. At \citet{Moran:2023}'s 0.2\% quoted precision in \ar, the uncertainty they would get in GJ 486's \rstar \ and \teff \ would be dominated by the 2.3\% uncertainty in \mstar, but we would still be concerned that such large systematics could bias the determination of \rhostar, \rstar \ and \teff. 

To ensure that light curve extraction systematics were not important for our analysis of \thisstar, we repeated the analysis described in \S \ref{sec:methods}, only swapping our Tiberius-extracted light curve with the Kempton-extracted (timestamp-corrected) light curve and found that the differences in \ar, \rhostar, \rstar, \teff, $e$, \rplanet, and \fave \ were all below $0.4\sigma$.
The only parameters that differed by more than $1\sigma$ were \rplanet/\rstar \ ($2.4\sigma$), the MIRI limb darkening coefficients ($1.2\sigma$), and the secondary eclipse time ($1.2\sigma$). We also note that \citet{Moran:2023}'s 0.2\% \ar \ uncertainty implies a \rhostar \ uncertainty that is 10$\times$ smaller than expected from equation \ref{eq:sigma_rhostar}.

Although we show in \S \ref{sec:discussion} that systematics in ground-based data are not relevant for our analysis of \thisstar, in general, it could be a significant factor and care must be taken to ensure that other sources of systematics (due to, e.g., airmass, atmospheric variability) are properly accounted for.

While negligible for \thisstar, there are other caveats that one should be aware of when applying this method more generally. For very rapidly rotating stars, the oblateness means the inferred radius is dependent on the star-planet orientation. This is negligible in the case of \thisstar \ but may be relevant for rapidly rotating stars. 

Other potential sources of systematic error in measuring \rhostar \ are detailed in \citet{Kipping:2014} and \citet{Eastman:2023}, but generally, one should be careful when applying this method to stars with light from another source that is unaccounted for, a nonneglibible planet mass, grazing transits, or non-Keplerian motion.

\section{Discussion}
\label{sec:discussion}

The ability to accurately measure the radii of M stars to such a degree is uncommon, having only been achieved for a small number of other low-mass targets. As \thisstar \ is an archetype for super-Earths around M dwarfs, precise constraints are invaluable and contribute to a deeper understanding of habitability, planet formation, and stellar evolution for all other M dwarf planetary systems. 

Our method applies to any high S/N transit with a low-mass host where the eccentricity can be measured very precisely. We identified six additional planets in the JWST archive with primary and secondary eclipses that should allow us to precisely measure this eccentricity as summarized in Table \ref{tab:observations}. 

The expected precision in \rhostar \ can be computed with the equation from  \citet{Stevens:2018},

\begin{equation}
\label{eq:sigma_rhostar} 
\frac{\sigma_{\rhostar}}{\rhostar} \approx \left(\frac{27}{2Q^2\theta}\right)^{1/2},
\end{equation}

\noindent where $Q \equiv \sqrt{\Gamma \tfwhm}\delta/\sigma$, $\theta \equiv \tau/\tfwhm$, $\sigma$ is the point measurement error at the data sampling rate $\Gamma$, $\delta$ is the transit depth, $\tau$ is the ingress/egress time, and \tfwhm \ is the duration of the transit from mid-ingress to mid-egress. This calculation ignores the contribution from limb darkening and eccentricity and should be treated as a rough guide, but for reference, our predicted 1.5\% precision of \rhostar \ for \thisstar \ is reasonably close to the \gjrhostarerr\% precision achieved above. 

While we included ground-based transits from 2012 and radial velocities (RVs) for our fit of \thisstar, it was not necessary for this method of determining a precise \rstar \ and \teff. We only did that to provide a new, definitive reference for \thisstar \ that incorporates the best knowledge of the planetary mass and period possible. We also ran a fit for \thisstar \ only using JWST data and found that the eccentricity, stellar radius, temperature, and density were constrained to a similar precision. The only change in precision was in the period and planet mass, which is expected. While the targets in Table \ref{tab:observations} also typically have a long baseline of ground-based observations and public RVs, we expect their contribution to \rhostar \ and $e$ to be similarly negligible.

We used the JWST Exposure Time Calculator to calculate the $\sigma$ as used in the equation for $Q$. The expected precision using equation \ref{eq:sigma_rhostar} is shown in Table \ref{tab:observations} as $\sigma_{\rhostar}/\rhostar$. The differences in expected precision for different light curves of the same object are entirely due to the expected difference in the S/N and ignore time sampling, covariance with limb darkening, and eccentricity uncertainty. 

\begin{table}[!h]
\centering
\setlength{\tabcolsep}{2pt}
\caption{JWST Archival Observations}
\begin{tabular}{lccc}
  \hline \hline
    Target name &      Event & $\sigma_{\rhostar}/\rhostar$ & Instrument \\
    \hline \\
         GJ 1132 & Eclipse     &     -- &             MIRI.LRS  \\
         GJ 1132 & Transit x 2 & 1.68\% &   NIRSPEC.BOTS+G395H \\
\hline
         GJ 1214 & Transit x 2 & 0.76\% &   NIRSPEC.BOTS+G395H  \\
         GJ 1214 &  PhaseC     & 1.53\% &             MIRI.LRS \\
\hline
          GJ 486 & Eclipse x 2 &     -- &             MIRI.LRS  \\
          GJ 486 & Transit x 2 & 3.25\% &   NIRSPEC.BOTS+G395H  \\
\hline
       LTT 1445A & Transit     & 1.94\% &   NIRSPEC.BOTS+G395H  \\
       LTT 1445A & Eclipse x 3 &     -- &             MIRI.LRS \\
\hline
        NGTS-10 &  PhaseC     & 0.34\% &   NIRSPEC.BOTS+PRISM \\
\hline
        WASP-43 &  PhaseC     & 0.18\% &   NIRSPEC.BOTS+G395H  \\
        WASP-43 &  PhaseC     & 1.28\% &             MIRI.LRS \\
\hline
        WASP-80 & Transit     & 0.44\% &             MIRI.LRS  \\
        WASP-80 & Eclipse     &     -- &             MIRI.LRS   \\
        WASP-80 & Transit     & 0.10\% & NIRCAM.GRISMR+F322W2 \\
        WASP-80 & Transit     & 0.16\% &  NIRCAM.GRISMR+F444W \\
        WASP-80 & Eclipse     &     -- & NIRCAM.GRISMR+F322W2 \\
        WASP-80 & Eclipse     &     -- &  NIRCAM.GRISMR+F444W \\
        WASP-80 & Transit     & 0.05\% &          NIRISS.SOSS \\
\hline\hline
\label{tab:observations}
\end{tabular}
\begin{flushleft}
\footnotesize{ \textbf{\textsc{NOTE:}}
 Note that the dashes in the table correspond to eclipses, which are used to determine the eccentricity and not \rhostar. Observations publicly available by Cycle 3 for $0.1 < \mstar/\msun < 0.7$ with at least one primary transit and secondary eclipse showing the predicted fractional uncertainty in \rhostar \ using equation \ref{eq:sigma_rhostar} and the JWST Exposure time calculator.
}
\end{flushleft}
\end{table}

Our method is immediately applicable to six additional stars currently available in the JWST archive, increasing the sample of precise stellar radii with which to calibrate future relations similar to \citet{Mann:2015} and is a valuable proof of concept to add many more. The number of possible low-mass targets that can be analyzed to achieve high-precision measurements similar to \thisstar \ using our method will continue to grow as JWST observations are taken for atmospheric science goals. In fact, in November 2023, the preliminary JWST Exoplanet working group recommended that JWST observe 15-20 rocky planets around M dwarfs, all of which should be good targets for this method.

While we have focused on low-mass host stars, the same basic technique applies to any star where we can measure the eccentricity using transits and occultations (e.g., from JWST), the density from the transit, the radius using a stellar mass, and a temperature from the bolometric flux. The only difference to account for is the stellar mass---the \citet{Mann:2019} relation used in our analysis does not apply for stars with a mass larger than 0.7 \msun. Instead, we could use a mass constraint from, e.g., evolutionary models.

\acknowledgements

We thank the anonymous referee for their insightful comments that improved the quality of our manuscript. We are grateful for discussions about systematic errors with Sarah Moran, Jacob Bean, Eliza Kempton, and Michael Zheng.

Work by A.S.M. and J.D.E. was funded by NASA ADAP 80NSSC19K1014. J.K.\ acknowledges financial support from Imperial College London through an Imperial College Research Fellowship grant.

This work is based in part on observations made with the NASA/ESA/CSA JWST. The data were obtained from the Mikulski Archive for Space Telescopes at the Space Telescope Science Institute, which is operated by the Association of Universities for Research in Astronomy, Inc., under NASA contract NAS 5-03127 for JWST. These observations are associated with program \#1803.

This research has made use of the Exoplanet Follow-up Observation Program website, which is operated by the California Institute of Technology, under contract with the National Aeronautics and Space Administration under the Exoplanet Exploration Program.

This research has made use of the NASA Exoplanet Archive, which is operated by the California Institute of Technology, under contract with the National Aeronautics and Space Administration under the Exoplanet Exploration Program.

This work has made use of data from the European Space Agency (ESA) mission
{\it Gaia} (\url{https://www.cosmos.esa.int/gaia}), processed by the {\it Gaia}
Data Processing and Analysis Consortium (DPAC,
\url{https://www.cosmos.esa.int/web/gaia/dpac/consortium}). Funding for the DPAC
has been provided by national institutions, in particular the institutions
participating in the {\it Gaia} Multilateral Agreement.

This publication makes use of data products from the Two Micron All Sky Survey, which is a joint project of the University of Massachusetts and the Infrared Processing and Analysis Center/California Institute of Technology, funded by the National Aeronautics and Space Administration and the National Science Foundation.

This publication makes use of data products from the Wide-field Infrared Survey Explorer, which is a joint project of the University of California, Los Angeles, and the Jet Propulsion Laboratory/California Institute of Technology, funded by the National Aeronautics and Space Administration.


\end{document}